\documentclass[12pt]{article}
\usepackage{amssymb}
\def\H{{\cal H}}
\def\B{{\sf b}}
\def\F{{\sf f}} 
\def\Z{{\bf Z}} 
\def\Z{\mathbb{Z}}
\begin{document}
\title{RR charges of D2-branes in the WZW model\\[1cm]}
\author{{\sc Anton Alekseev}\\[2mm] Institute for Theoretical Physics, 
Uppsala University,\\ Box 803, S--75108 Uppsala, Sweden\\[5mm]    
{\sc Volker Schomerus}\\[2mm]  
MPI f\"ur Gravitationsphysik, Albert-Einstein-Institut,\\ 
Am M\"uhlenberg 1, D--14476 Golm, Germany}

\date{June 2000}

\maketitle

\begin{abstract}
We consider the contribution of the $B$-field into the
RR charge of a spherical D2-brane. Extending a recent 
analysis of Taylor, we show that the boundary and bulk 
contributions do not cancel in general. Instead, they 
add up to an integer as observed by Stanciu. The 
general formula is applied to compute the RR charges 
of spherical D-branes of the $SU(2)$ WZW model at level 
$k$ and it shows that these RR charges are only defined 
modulo $k+2$. We support this claim by studying bound 
state formation of D0-branes using boundary conformal 
field theory. 
\end{abstract}

The issue of Ramond-Ramond (RR) charges in background fluxes
was raised by Bachas, Douglas and Schweigert \cite{BDSch}
(see also \cite{Paw}). 
They calculated the RR charges of spherical D2-branes 
which arise in the WZW model on the group $SU(2)$ \cite{ASch}, 
\cite{Gaw}, \cite{FFFSch}.
Two calculations, the first one based on the semiclassical considerations
and the second one using the exact 1-point functions in the boundary
WZW model, indicated that RR charges of D2-branes are irrational
and that their differences are not integral. This observation
is in contradiction with our intuition about the nature $U(1)$ 
charges and, hence, gives rise to an interesting puzzle which 
we discuss in this note. We shall see that there is a natural 
way to assign a $\Z_{k+2}$ valued charge to D2-branes. Even though this
charge is first motivated within the semi-classical approach, we 
conjecture that it contains important dynamical information. 
According to \cite{ARS}, a stack of D0-branes on a large $SU(2)$ 
is unstable against decay into spherical D2-branes. Our 
conjecture is that the integer RR charge of a D2 
brane counts the number of its constituent D0-branes and
that this number is conserved ({\it mod} $k+2$) in all physical 
processes. We provide some evidence for this proposal but it 
certainly has to undergo further tests.    
\medskip

Let us recall that in the case of flat D-branes the WZ coupling
of Ramond-Ramond fields $C= \sum_i c_i$ on the D-brane
world-volume is given by formula  \cite{flat},
\begin{equation} \label{Iflat}
I^{flat} \ = \ \int_V\,  C \wedge \exp(B+F)\ \ ,
\end{equation}
where $B$ is the bulk $B$-field, $F$ is the curvature
of the $A$-field on the brane, and $V$ is the world-volume
of the D-brane $D$. At this point it is convenient
to introduce the special notation $B_D = B+F$. We call this
combination a {\em boundary} $B$-field. Note that in contrast
to $B$ and $F$ the field $B_D$ is gauge invariant since it
defines the boundary conditions for the open string action.

Equation (\ref{Iflat}) implies the following expression for the
D0-brane RR charge of a D2-brane,
\begin{equation}  \label{Qflat}
Q_{RR}^{flat}\ =  \   \int_{D} B_D\ \ .
\end{equation}
In the case of a contractible D2-brane, the right hand side
of equation (\ref{Qflat}) is quantized in the units of $2\pi$.
Indeed, the expression for the RR charge is proportional
to the flux $\Phi =  2\pi \int_D B_D$ of the field $B_D$
through the D-brane. Only the D-branes with integral
flux, $\Phi=2\pi n$ with $n$ integer, define consistent 
boundary conditions for open strings. Thus, the flux quantization
condition implies that $Q_{RR}^{flat}$ is an integer.

In the case of curved D-branes  formulas (\ref{Iflat})
and (\ref{Qflat}) receive corrections \cite{MM},
\begin{equation} \label{I}
I \ = \ \int_V \, C \wedge \exp(B_D + \frac{1}{2} d) \hat{A}(TD)
\, \frac{1}{f^* \sqrt{\hat{A}(TS)}}\ \  .
\end{equation}
Here $d$ is the degree two class defining the {\em Spin}$^c$
structure of $D$, $\hat{A}(TD)$ is the $A$-roof genus of the
D-brane, $S$ is the space-time, and $f: D \rightarrow S$
is the embedding of the D-brane to the space-time.
In this paper we only consider  situations when 
the tangent bundle $TS$ is trivial. Then, formula
(\ref{I}) simplifies since $\hat{A}(TS)=1$ and 
$d = c_1(TD)$,
\begin{equation} \label{Ifinal}
I \ = \ \int_V \, C \wedge \exp(B_D) {\rm Td}(TD)\ \ ,
\end{equation}
where ${\rm Td}(TD)$ is the Todd genus of $D$. This leads
to the following formula for the RR charge,
\begin{equation} \label{Qfinal}
Q_{RR} \ = \ \int_D \exp(B_D) {\rm Td}(TD).
\end{equation}
By Atiyah-Singer index theorem, the right hand side
is an index of the $Spin^c$ Dirac operator on the D-brane
twisted by the line bundle with the first Chern class
represented by the 2-form $B_D$.
Hence, $Q_{RR}$ is an integer. In the case of D2-branes formula
(\ref{Qfinal}) simplifies,
\begin{equation} \label{Q}
Q_{RR} \ = \ \int_D  B_D +  \frac{1}{2} c_1(TD) \ =
\ \frac{\Phi}{2\pi} + \frac{1}{2} c_1(TD)\ \ .
\end{equation}
\medskip

In  applying  formula (\ref{Q}) to the discussion of  
D2-branes in the $SU(2)$ WZW model we begin with the simpler
limit in which the level $k$ is sent to infinity. In this case
one is only interested in the small neighborhood of the group
unit and one can replace the sphere $S^3 \cong SU(2)$ by the
Euclidean space $R^3$. The D-branes are 2-spheres centered
at the origin. They carry the flux of the Neveu-Schwarz (NS) 
$B$-field proportional
to their radius $\Phi(r) = 2\pi r$. Only the spheres with 
integral radius give rise to consistent open string boundary
conditions, $\Phi(n) = 2\pi n$. In our case, $c_1(TD)=2$,
and we obtain $Q_{RR}(n) = n+1$. Note that in the definition of the
Chern class $c_1(TD)$ we use the complex structure on $D$
such that the metric defined by $B_D$ be positive definite.
\smallskip

\def\bz{{\bar z}}
\def\nn{\nonumber} 
\def\o{{\otimes}}
Let us briefly recall how this conclusion can 
be obtained from a CFT calculation of the RR charge \cite{BDSch}. Suppose 
we are dealing with the supersymmetric WZW theory at level $k+2$. 
This model contains currents $J^a$ satisfying the 
relations of a level $k+2$ affine Kac Moody algebra along with 
a multiplet of free fermionic fields $\psi^a$ in the adjoint 
representation of $su(2)$. It is well known that one can 
introduce new bosonic currents 
$$ J^a_\B \ := \ J^a + \frac{i}{k} f^{a}_{\ bc} \psi^b \psi^c $$ 
which obey again the commutation relations of a current algebra 
but now the level is shifted to $k$. The fermionic fields $\psi^a$ 
commute with the new currents. This means that the theory splits 
into a product of a level $k$ WZW model and a theory of three 
free fermionic fields. 

We want to impose gluing conditions $J^a(z) = \bar J^a(\bz)$ 
and $\psi^a (z) = \pm \bar \psi^a(\bz)$ along the boundary $z = \bz$. 
This obviously implies $J^a_\B (z) = \bar J^a_\B (\bz)$. Hence, the 
boundary states of the theories we are studying, factorize 
into two well known contributions,  
$$ |\alpha,\pm \rangle^{\rm susy}  \ = \ |\alpha\rangle \o |\F,\pm \rangle  
   \ \ \ \mbox{ for } \ \ \ \alpha = 0,\frac12, \dots, \frac{k}{2}
\ \ , $$
where $|\alpha \rangle$ is one of the boundary states for a WZW-model
at level $k$ and $|\F, \pm \rangle$ denote the familiar fermionic 
boundary states for (anti-)branes in flat space. 

The boundary states of the level $k$ WZW-model can be characterized 
by the quantities: 
$$ \langle j | \alpha \rangle \ = \ \frac{S_{\alpha j}}{\sqrt{S_{0 j}}} 
\ = \ \left(\frac{2}{k+2}\right)^{1/4} \ \frac{\sin \pi 
         \frac{(2\alpha +1)(2j+1)}{k+2}}
           {\left(\sin \pi \frac{2j+1}{k+2}\right)^{1/2}} \ \ . $$
Here $|j\rangle$ denotes a spin $j$ primary field for the currents 
$J_\B$. 

For a discussion of the fermionic states we refer the reader to 
the standard literature (see e.g.\ \cite{Pol}). Let us only remark 
that these states $|\F,\pm \rangle = |{\rm NSNS}\rangle \pm |{\rm RR}
\rangle$ contain contributions from the NSNS and the RR sector. The 
choice of the sign $\pm$ distinguishes between branes and anti-branes.

We obtain the RR charge of the brane by computing $\langle \phi |
\alpha,+\rangle^{\rm SUSY}$ for an appropriate state $ |\phi\rangle$ 
from the (RR) sector of the bulk theory. $|\phi\rangle$ is again a
product state containing contributions from the RR ground states
of the three free fermions and of the $(j,\bar j)=(1/2,1/2)$ 
representation for the currents $J^a$. Following \cite{BDSch},
 we insert this into  the formulas above to  obtain
$$ Q_{RR}(n) \ = \ \lim_{k \rightarrow \infty} \, 
                 \frac{\langle 1/2| \alpha,+\rangle^{\rm susy}} 
                        {\langle 1/2| 0,+ \rangle^{\rm susy}} 
             \ = \ n+1 \ .
$$      
Hence we reproduce precisely the RR charges from the previous 
discussion based on formula (\ref{Q}). 
\bigskip

Now we turn to the case of finite $k$ where the bulk
curvature of $S^3$ cannot be ignored. It leads to
a non-vanishing field strength $H$ of the bulk NS
$B$-field. Recently, Taylor observed that for
homotopically trivial D2-branes there is an extra
contribution to the RR charge of the D2-brane
coming from the integral of the $H$-field. More precisely,
for a D2-brane $D$ which can be contracted to a point along
a 3-manifold $\Gamma$ with $\partial \Gamma = D$, the 
contribution of the $H$-fields to the RR charge of the
brane is given by (see (4.1) and (4.4) of \cite{Tay}),
\begin{equation} \label{Tay}
\Delta Q_{RR}\  = \ - \int_\Gamma H\ \ .
\end{equation}
The total RR charge of the D2-brane reads,
\begin{equation} \label{answer}
Q_{RR} \ =\ 
\int_D B_D -  \int_\Gamma H + \frac{1}{2} c_1(TD)\ \ .
\end{equation}
This agrees with the results of Stanciu \cite{Sta}
except from the correction by $\frac12c_1(TD)$. We 
can always choose the bulk $B$-field such that
on $\Gamma$ one obtains,  
\begin{equation}  \label{HdB}
H=dB.
\end{equation}
Then, the contributions of $B$ and $H$-fields to
$Q_{RR}$ cancel each other and we obtain,
\begin{equation} \label{QF}
Q_{RR} \ = \ \int_D F + \frac{1}{2} c_1(TD)\ \ .
\end{equation}
We extrapolate the result of \cite{MM} by requiring
that $c_1(TD)$ is obtained using the (almost) complex structure
on $D$
such that the metric defined by $F$ be positive definite.
The extra contribution $\frac{1}{2} c_1(TD)$ is the
correction to Taylor's formula.  

In the presence of  $H$-field the flux integrality
condition imposed on the brane is modified \cite{Klim}
and reads 
(see {\em e.g.} equation (9) in \cite{ASch}),
\begin{equation}  \label{PhiH}
\Phi \ = \ 2\pi \left( \int_D B_D - \int_\Gamma H \right) \ =\  
2\pi \int_D F \ = \ 2\pi n
\end{equation}
with $n$ integer. For the D2-branes of the WZW model 
the combination of (\ref{QF}) and (\ref{PhiH}) yields, 
\begin{equation}   \label{Qn}
Q_{RR}(n) \ = \ \frac{\Phi}{2\pi} + \frac{1}{2} c_1(TD) \ = \ n+1\ ,
\end{equation}
which is exactly the same formula as we had in the case
$k \rightarrow \infty$. In the case of finite $k$ the difference
is that there is only a finite number of spherical D2-branes
corresponding to $n= 1, \dots, (k-1)$. These branes carry RR
charges equal to $Q_{RR} = 2, \dots, k$, respectively.
We provide explicit expressions for $H$, $B$, $B_D$ and $F$
for the WZW model in Appendix.
We conclude that in Lagrangian approach the self-consistent
boundary conditions for open strings necessarily lead to
integral RR charges, as expected on  general grounds.

There is a new feature of the RR charges on curved backgrounds
which follows from our consideration. 
Since the $F$-field is not a gauge invariant object,
different choices of $B$ and $F$ corresponding to the
same value of $B_D = B+F$ may give rise to different
values of $Q_{RR}$.
In the case of the $SU(2)$
WZW model the spherical D-branes can be contracted either to
the group unit $e$ along the 3-ball $\Gamma$ 
or to the opposite pole $(-e)$ of $S^3$ along the 3-ball $\Gamma'$.
The picture viewed from $(-e)$ looks exactly the same way as
from $e$ with the exception that the value of the RR charge
changes the sign because the 2-form $F$ changes sign
under the transformation $n \rightarrow (k-n)$
together with $\Gamma \rightarrow \Gamma'$, and, hence,
one should choose the opposite complex structure on $D$
such that $c_1(K_D)$ also changes sign. That is,
we obtain another value of the RR charges of the same
spherical D2-branes,
\begin{equation}  \label{Qn'}
Q'_{RR}(n) \ = \ - ( (k-n) +1) = (n+1) - (k+2)\ \ .
\end{equation}
Here the shift by $k$ is due to the change of the 3-ball
$\Gamma$ to $\Gamma'$ and the extra shift by 2 is due to
the change of sign on the canonical bundle when we
shift from $e$ to $(-e)$. We conclude that the RR charge 
is only well defined modulo $(k+2)$. This is a novel 
feature which arises because the $H$-field belongs to
a nontrivial cohomology class.
\bigskip

We would like to confirm the conclusions of the Lagrangian 
approach by computing $Q_{RR}$ in the CFT picture. 
The standard prescription suggests computing RR charges
from overlaps of  closed string states with the boundary state.
Unfortunately,  in case of finite level $k$, such overlaps typically 
change along the RG-trajectories which are generated by boundary 
fields. That is, RR charges will not be conserved in physical
processes such as formation of bound states of D-branes.

Instead, we shall conjecture that formulas (\ref{Qn}-\ref{Qn'})
give an RG-invariant  definition of RR charges 
which  are defined only {\it mod\/}  $k+2$. 
We test this conjecture by arguing that an anti-brane at $-e$ can 
form as a bound state of $k+1$ branes at the origin $e$.  
In this process the state with RR charge $k+1$ evolves into
the state with RR charge $-1$.

\medskip

We assume that the boundary state $|0,+\rangle^{\rm susy}$
describes a brane at $e$. Then,  the state $|k/2,+\rangle^{\rm susy}$
corresponds to an anti-brane at $-e$. Indeed, we can translate 
branes on the 3-sphere by acting with the currents 
$J^a = J^a_\B + J^a_\F$ where $ J^a_\F := - \frac{i}{k} f^{a}_{\ bc} 
\psi^b \psi^c$. The finite translation which moves D0-branes 
from $e$ to $-e$ corresponds to the element $-e$ in the group. 
One can check easily that the first factor $|0\rangle $ in $|0
\rangle^{\rm susy} = |0\rangle \otimes |\F,\pm \rangle$ is 
shifted to $|k/2\rangle$ (see e.g.\ \cite{ReSc2}). To see the 
effect of translations on the fermionic factor one should recall 
that the RR sector contains states with half-integer spins 
$(j_\F,\bar j_\F) \in (\frac12 + \Z) \times (\frac12 + \Z)$ 
while NSNS states are built up from integer spin only. Since 
the element $-e$ is represented by $\pm 1$ depending on whether
it acts on states with integer or half integer spin, we 
conclude that $|\F,+\rangle = |{\rm NSNS}\rangle + |{\rm RR}
\rangle  $ gets mapped into $|\F,-\rangle = |{\rm NSNS}\rangle - 
|{\rm RR}\rangle$ by the finite translation with $-e \in SU(2)$. 
Hence, $|k/2,-\rangle ^{\rm susy}$ is the boundary state of 
a brane at $-e$ which means that $|k/2,+\rangle ^{\rm susy}$ 
describes an anti-brane at $-e$. 
\medskip

Now let us look at a stack of $k+1$ D0-branes at the origin 
$e$. Following \cite{ARS}, this stack is expected to be unstable
against perturbation with $S_a J^a(x)$ where $S_a$ are $k+1$ 
dimensional representation matrices of $su(2)$. This perturbation
triggers a decay into some bound state. It is difficult to obtain
exact statements on the  nature of this bound state but there 
exists a heuristic 
rule that seems to produce correct results in condensed matter physics
\cite{AffLud} where it is used to identify the low temperature 
fixed point of the Kondo model. 

To be more specific, let us 
consider open strings that stretch between the stack of $k+1$ branes 
at $e$ and a single anti-brane at $e$. The corresponding states 
are taken from the space $V_{k+1} \o \H_{e^+,e^-}$ where $V_{k+1}$ 
is a $k+1$-dimensional vector space and $\H_{e^+,e^-} = \H^{0}_{k} 
\o \H^\F_-$ is the product of the vacuum module for the current 
algebra at level $k$ with the fermionic space $\H^\F_-$. The 
latter carries an action of the level $2$ Kac-Moody algebra
generated by $J^a_{\F}$ and decomposes into a direct 
sum of $\H_2^0$ (NS-sector) and $\H_2^1$ (R-sector). The 
appearance of the vacuum sector gives rise to the tachyonic 
mode in the brane--anti-brane system. 

On our space $V_{k+1} \o \H_{e^+,e^-}$ there are commuting 
actions of the $k+1$-dimensional matrices $S_a$ on the first 
tensor factor $V_{k+1}$ and of the currents $J^a$. 
These two symmetries are broken by the perturbation with $S_a 
J^a$. Experience with similar couplings of spin and angular 
momentum in standard atomic physics  leads us to expect that 
the sum $S_a + J^a$ has a good chance to be preserved. 
Note that the generators $S_a + J^a_n$ which are obtained 
by shifting all modes $J^a_n$ of the current $J^a$ with the 
same constant $S_a$, satisfy the relations of a current algebra 
at level k+2. The rule about the conservation 
of $S_a + J^a$ has been tested through CFT investigations of the 
Kondo model \cite{AffLud} and it does correctly reproduce the results 
of \cite{ARS} at large level $k$. We shall use it here to determine 
the open string spectrum of our configuration. Note that $\H_{e^+,e^-}$ 
carries a tensor product of the spin $j_\F = 0, \frac12$ representations 
of the fermionic currents with the spin $j_\B = 0$ representation for 
$J^a_\B$. These add up to spin $j = 0, \frac12$ representations 
of the level $k+2$ currents $J^a$. If we add the spin 
$s=k/2$ representation on $V_{k+1}$ we end up with a spins 
$k/2, k/2 \pm 1/2$. Thus we have shown that the representation 
of current algebra on  $V_{k+1} \o \H_{e^+,e^-}$ generated by $S_a 
+ J^a_n$ decomposes into representations of spin $k/2,k/2 
\pm 1/2$.  

Our aim is now to compare this to the spectrum of open strings
stretching between a single anti-brane at $-e$ (the conjectured 
bound state of the original stack) and the anti-brane at $e$. 
Here we employ our previous claim that $|k/2,+\rangle$ describes
the anti-brane at $-e$. This implies that the wave functions of 
open strings stretching between the anti-branes at the poles 
are taken from the space $\H_k^{k/2} \otimes \H^\F_-$. Under 
the action of $J^a$ the latter decomposes into irreducibles
of spin $j = k/2, k/2 \pm 1/2$. This are precisely the spin 
values we found for $S_a + J^a$ in the initial configuration. 

The analysis supports or claim that $k+1$ D0-branes at $e$ 
decay into a single anti-brane at $-e$. If we want the 
RR charge to be conserved, we are forced to identify the 
charges of these configurations. In this sense the CFT 
considerations seem to confirm our proposal that the 
RR charge is defined only {\it mod\/} $k+2$.          
\bigskip\bigskip

\noindent
{\bf Acknowledgements:} We would like to thank C.\ Bachas, 
I.\ Brunner, C.S.\ Chu, A.\ Morozov, A.\ Recknagel and
Chr.\ R\"omelsberger for interesting discussions.

\section*{Appendix}

Here we collect explicit formulas for the fields $H$, $B$ and $B_0$
in the case of the $SU(2)$ WZW model.

We recall \cite{ASch} that in this case the field $B_D$
which determines the open string boundary conditions on the 
D-brane is given
by formula,
\begin{equation}
B_D\ = \ \frac{k}{8\pi} {\rm Tr} \, 
\left( dgg^{-1} \, \frac{1+Ad(g)}{1-Ad(g)} \, dg g^{-1} \right)\ \ ,
\end{equation}
where $k$ is the level on the WZW model. In terms of the Euler angles
$\psi, \theta$ and $\phi$ this expression can be rewritten as
$$
B_D\ = \ \frac{k}{2\pi}  \sin(2\psi) \sin(\theta) d\theta d\phi\ \ .
$$
The $H$-field is simply equal to the WZW form,
\begin{equation}  \label{H}
H\ = \ \frac{k}{12\pi} {\rm Tr} (dgg^{-1})^3 = 
\frac{2k}{\pi} \sin^2(\psi) \sin(\theta) d\psi d\theta d\phi\ \ .
\end{equation}
In terms of the Euler angles, one can solve equation
$H=dB$ with (see equation (2.3)
of \cite{BDSch}),
$$
B\ = \ \frac{k}{\pi}  \left( \frac{\sin(2\psi)}{2} 
- \psi \right) \sin(\theta) d\theta d\phi\ \ .
$$
Finally, the field $F$ is obtained as a difference of $B_D$ and $B$,
$$
F\ = \ \frac{k}{\pi} \psi \sin(\theta) d \theta d \phi\ \ ,
$$
with integral over D given by formula,
\begin{equation}
\int_D F \ = \ 2  k \psi\ \ .
\end{equation}
For integral D-branes of the WZW model $\psi_n = \pi n/k$ which yields
the contribution to the RR charge equal to $n$.

\end{document}